\documentclass{ws-ijgmmp}
\begin{document}

\markboth{Piyali Bhar and Pramit Rej}
{Charged gravstar model in $f(T)$ gravity admitting conformal motion}

%
\catchline{}{}{}{}{}
%

\title{Charged gravastar model in $f(T)$ gravity admitting conformal motion}

\author{Piyali Bhar}
\address{Department of Mathematics, Government General Degree College, Singur,
Hooghly,\\ West Bengal-712409, India\\
\email{piyalibhar90@gmail.com}}

\author{Pramit Rej}
\address{Department of
Mathematics, Sarat Centenary College, Dhaniakhali, Hooghly, \\West Bengal 712 302, India\\
\email{pramitrej@gmail.com}}

\maketitle

\begin{history}
\received{(Day Month Year)}
\revised{(Day Month Year)}
\end{history}

\begin{abstract}
In present paper model of charged gravastar under $f(T)$ modified gravity is obtained. The model has been explored by taking the diagonal tetrad field of static spacetime together with electric charge. To solve the Einstein-Maxwell field equations, along with $f(T)$ gravity, we assume the existence of a conformal Killing vector which relates between geometry and matter through the Einstein Maxwell field equations by an inheritance symmetry. We study several cases of interest to explore physically valid features of the solutions. Some physical properties of the model are discussed and we match our interior spacetime to the exterior Reissner-Nordstr\"{o}m spacetime in presence of thin shell.
\end{abstract}











\section{Introduction}
In modern cosmology, one of the most important problems is to deal with the dark energy issue which causes the accelerating expansion of the Universe. This phenomena has been confirmed by numerous observations of large scale structure\cite{1,2} and measurements of the cosmic microwave
background(CMB) anisotropy \cite{3,4}. The source that drives this cosmic acceleration is termed as `dark energy' and it possess positive
energy density but negative pressure. It is well known that this form of energy is acting as a repulsive gravitational force so that in General Relativity (GR) one needs to consider a further non-standard fluid with a negative pressure to justify this accelerated scenario. To explain the
accelerating phase of the Universe, several modified theories on gravity have been proposed. A few of them are $f(R)$ gravity, $f(T)$ gravity and $f(R,T)$ gravity, EGB gravity, $f(G)$ gravity, $f(G,\,T)$ gravity etc. which have received more attention in recent decades.\par
Since last few years, $f(T)$ gravity drew the attentions to the researchers in the field of cosmology due to the possibility to explain by it the accelerated expansion of the cosmic Hubble fluid. In addition, astrophysical studies related with compact objects as black holes has been considered. In $f(T)$ gravity theory, an arbitrary analytic function of the torsion scalar $T$ is used in place of the gravitational part in the standard Einstein-Hilbert action. The advantage of using the $f(T)$ theory of gravity over the $f(R)$ theory of gravity is that in case of $f(T)$ theory of gravity the field equations provides the differential equations of order $2$ whereas in the latter case, the field equations turns out in fourth order differential equations, which is difficult to handle \cite{12}. One can remember that Einstein himself proposed the idea of the Teleparallel Theory
in $1928$ to unify gravity and electromagnetism and this theory is now
formulated as higher gauge theory \cite{7}. Though this problem of unification can not be solved in the new theory, after many years the theory becomes much popular as an alternative theory of GR , and this is known as teleparallel equivalent of GR (TEGR).\par
Ganiou \cite{gan} investigated the structures of neutron stars under the strong magnetic
field in the framework of $f(T)$ gravity.  Ahmed et al. \cite{ahmed} studied the accretion process for fluids flowing near a black hole in the context of $f(T)$ teleparallel gravity. Jamil et al \cite{jamil} examined the interacting dark energy model in $f(T)$ cosmology by assuming dark energy as a perfect fluid and choose a specific cosmologically viable form $f(T) = \beta \sqrt{T}$. The stability
and phase space behavior of the parameters of the scalar field by choosing an exponential
potential and cosmologically viable form of $f(T)$ were investigated by Jamil et al \cite{jam} and the authors found that the dynamical system
of equations admit two unstable critical points and they conclude that no attractor solutions exist in the
cosmology. Strange Stars in $f(T)$ Gravity with MIT bag Model in the background of KB metric with electric field was obtain by Abbas et al. \cite{abbas}. By assuming conformal killing equation together with $f(T)$ gravity model of compact star was obtained by Das et al. \cite{das}. $f(T)$ theory has also been used to obtain the model of Black hole and compact objects in 3-dimensions. On the basis of this fact, the BTZ black hole model has been derived in $f(T)$ theory of gravity \cite{zheng}. Ulhoa and Rocha \cite{ur} investigated the structures of neutron stars using a a perfect fluid model within the context of the teleparallel equivalent to general relativity (TEGR). The author used numerical computations (based on the RNS code) to find the relationship between the angular momentum of the field and the angular momentum of the source. They also derived the regime where linear relation between gravitational angular momentum and moment of inertia is valid. Nashed and Capozziello \cite{nc} derived magnetic black hole solutions by employing a general gauge potential within the framework of TEGR. They calculated the torsion and curvature invariants to discuss the singularities and also derived the conserved quantities using the Einstein-Cartan geometry to realize the physics of the constants appearing in the solutions. Capozziello and Nashed \cite{cn} derived exact charged d-dimensional black hole solutions for quadratic type teleparallel equivalent gravity in terms of torsion scalar. They calculated the electromagnetic Maxwell field and proved that the d-dimensional black hole solutions coincide with the previous obtained result of  Awad et al \cite{awad}. The authors also investigated the thermodynamical properties of the solutions by calculating the entropy, the Hawking temperature, the heat capacity, and other physical quantities. Very recently, Nashed and Capozziello \cite{nc1} derived a charged non-vacuum solution for a physically symmetric tetrad field with two unknown functions of radial coordinate within the framework of Teleparallel Gravity theory and studied the stability of the model through various conditions. Some earlier work on the modelling of compact star in $f(T)$ gravity are found in the references \cite{12,20,21,22,24}.


\par
A gravastar is an astrophysical object which is hypothesized as an alternative to the black
hole theory proposed by Mazur and Mottola \cite{maz,maz1}.
The model proposed by Mazur and Mottola \cite{maz,maz1} corresponding to the exterior schwarzschild geometry, is a static spherically symmetry perfect fluid model and it has three different regions. In each region of the gravastar the relation between isotropic pressure $p$ and matter density $\rho$ is described as follows:
\begin{enumerate}
\item Interior region : ~~~~$0\leq r_1 < r,$ ~~~~~$p+\rho=0$
\item Thin Shell :~~~ $r_1< r< r_2 ,$ ~~~~~$p=\rho$
\item Exterior :~~~ $r_2< r$,~~~~ $p=\rho=0$
\end{enumerate}
Where $p$ represents the isotropic pressure and $\rho$ is the energy density of the perfect fluid sphere.
Lobo and Garattini \cite{lobo} explored the gravastar solutions in the context of
noncommutative geometry along with their physical properties and characteristics.
Cattoen et al.\cite{viser} proposed that gravastars cannot be perfect fluids, it has anisotropic pressures in the `crust' by considering
the usual TOV equation for static solutions with negative central pressure.
In the frame-work of the
Mazur-Mottola \cite{maz} model, Usmani et al.\cite{usmani} proposed a new charged
gravastar model by assuming conformal motion corresponding to the exterior Reissner-Nordstr\"{o}m line-element and it was generalized by Bhar \cite{bhar14} in higher dimensional spacetime. Rahaman et al. \cite{rah1,rah2} obtained a model of both uncharged and charged gravastar
in ($2$+$1$) dimensional anti-de Sitter spacetime and its exterior spacetime is the BTZ spacetime.


Our present paper is inspired by all of these previous work done. Following the model of gravastar proposed by Mazur-Mottola \cite{maz}, in the present paper we obtain a model of charged gravastar in the background of conformal symmetry of the space-time in $f(T)$ theory of gravity. The present paper is decorated as follows: In Sect.~\ref{sec2} we give a brief of Einstein-Maxwell field equation in $f(T)$ gravity. In next section, the outline of the conformal killing vectors is described. The field equations in $f(T)$ gravity are solved in Sect.~\ref{sec4}. Some physical properties of the model is discussed in the next section and some concluding remarks are given in Sect.~\ref{sec6}.

\section{Interior spacetime and Einstein-Maxwell Equation in $f(T)$ theory}\label{sec2}
To describe the interior of a static spherically symmetry spacetime, we take the line element in standard form as,
\begin{equation}\label{line}
ds^{2}=e^{\nu(r)}dt^{2}-e^{\lambda(r)}dr^{2}-r^{2}(d\theta^2+\sin^2\theta d\phi^2).
\end{equation}
where $\nu$ and $\lambda$ are the two unknown functions of the radial co-ordinate `r'. In our present discussion we use the geometric unit i.e., $G=1=c$.\par
Now by the matrix transformation called
tetrad, This line element can be converted to a Minkowskian space as follows :
\begin{eqnarray}
ds^2&=&g_{\mu \nu}dx^{\mu}dx^{\nu}=\eta_{ij}\theta^i\theta^j,\\
dx^{\mu}&=&e_i^{\mu}\theta^i,~\theta^i=e^i_{\mu}dx^{\mu}.  \label{tet}
\end{eqnarray}
where
$\eta_{ij}=diag(1,-1,-1,-1)$ and
$e_{i}^{\mu}e^{i}_{\nu}=\delta^{\mu}_{\nu}$. The determinant of the metric $g$ is connected to the determinant of the tetrad
$e=\sqrt{-g}=det(e^{i}_{\mu})$.\\
The components of the torsion and the contorsion  are respectively defined as \cite{bb}
\begin{eqnarray}
T^{\sigma}_{\mu\nu}&=&\Gamma^{\sigma}_{\nu\mu}-\Gamma^{\sigma}_{\mu\nu}=e^{\sigma}_{i}\left(\delta_{\mu}e^{i}_{\nu}-\delta_{\nu}e^{i}_{\mu}\right),\\
K^{\mu\nu}_{\sigma}&=&-\frac{1}{2}\left(T^{\mu\nu}_{~~\sigma}-T^{\nu\mu}_{~~\sigma}-T^{\mu\nu}_{\sigma}\right),
\end{eqnarray}
and the components of the tensor $S_{\sigma}^{\mu \nu}$ is described as,
\begin{equation}
S^{\mu\nu}_{\sigma}=\frac{1}{2}\left(K^{\mu\nu}_{~~\sigma}+\delta^{\mu}_{\sigma}T^{\beta\nu}_{~~\beta}-\delta^{\nu}_{\sigma}T^{\beta\mu}_{~~\beta}\right),
\end{equation}
The torsion scalar is obtained from torsion and contorsion scalar are as follows:
\begin{equation}
T=S^{~~\mu\nu}_{\sigma}T^{~~\sigma}_{\mu\nu}.
\end{equation}
Now the action of $f(T)$ theory is given as:
\begin{equation}
S[e^i_{\mu},\phi_A]=\int{d^4x~e~\left[\frac{1}{16\pi}f(T)+\mathcal{L}_{matter}(\phi_A)\right]},
\end{equation}
Here, $\phi_A$ are matter fields and $f(T)$ is an arbitrary analytic function of the torsion scalar $T$.


Now varying the action with respect to the tetrads, the field equations of $f(T)$ gravity can be obtained as \cite{li}
\begin{equation}
S^{~~\nu\rho}_{\mu}\partial_{\rho}T f_{TT}+\Big[e^{-1}e^i_{\mu}\partial_{\rho}(e e^{~~\alpha}_iS^{~~\nu \rho}_{\alpha})\Big]f_{T}
+\frac{1}{4}\delta^{\nu}_{\mu}f=4\pi\Upsilon^{\nu}_{\mu},
\end{equation}
where
\begin{equation}
f_T=\frac{\partial
f}{\partial T}~~~~f_{TT}=\frac{\partial^2f}{\partial T^2},
\end{equation}
and $\Upsilon^{\nu}_{\mu}$ represents the energy momentum tensor of the underlying fluid.\\
Let us consider a matter distribution consisting of a charged fluid which is locally isotropic, i.e., radial pressure $(p_r)$ and the transverse pressure $(p_t)$ are equal $(p_r=p_t=p)$. Then $\Upsilon^{\nu}_{\mu}$ has two components one is for matter ($\mathcal{T}^{\nu}_{\mu}$) and another is for charge ($\mathcal{E}^{\nu}_{\mu}$). Therefore we can write,
\begin{equation}
\Upsilon^{\nu}_{\mu}=\mathcal{T}^{\nu}_{\mu}+\mathcal{E}^{\nu}_{\mu},
\end{equation}
The energy momentum tensor for the matter and charge are respectively given by,
\begin{eqnarray}
\mathcal{T}^{\nu}_{\mu}&=&(p+\rho)u^{\nu}u_{\mu}-p\delta^{\nu}_{\mu},\\
\mathcal{E}^{\nu}_{\mu}&=&\frac{1}{4\pi}\left(\frac{1}{4}\delta^{\nu}_{\mu}F_{\alpha\beta}F^{\alpha\beta}-F^{\nu \alpha}F_{\mu \alpha} \right).\label{ef}
\end{eqnarray}
where $u^i$ is the four-velocity and $\rho$ is the matter density of the fluid. In eqn. (\ref{ef}) $F_{\alpha \beta}$ is the electromagnetic field tensor and it can be defined by
\begin{equation}\label{14}
F_{\alpha\beta}=\frac{\partial A_{\beta}}{\partial A_{\alpha}}-\frac{\partial A_{\alpha}}{\partial A_{\beta}},
\end{equation}
where $F_{\alpha\beta}$ satisfy the Maxwell equations,
\begin{eqnarray}
F_{\alpha\gamma,\beta}+F_{\gamma \beta,\alpha}+F_{\beta \alpha, \gamma}&=&0,\\
\text{and}~~~~~~~\frac{\partial}{\partial x^{\gamma}}(\sqrt{-g}~F^{\alpha \gamma})&=&-4\pi\sqrt{-g}~J^{\alpha}.
\end{eqnarray}
where $g=det(g_{ij})$ and $A_\alpha=(\phi(r),0,0,0)$ is the four-potential and $J^{\alpha}$ is the four current vector defined by $J^{\alpha}=\sigma u^{\alpha}.$
$\sigma(r)$ is the electric charge density.
\par
We re-write the line element (\ref{line}) into the invariant form under the Lorentz transformations by defining the tetrad matrix (\ref{tet}) as
\begin{equation}
e^{i}_{\mu}=\begin{pmatrix}
              e^{\frac{\nu}{2}} & 0 & 0 & 0 \\
              0& e^{\frac{\lambda}{2}} & 0 & 0 \\
              0 & 0 & r& 0 \\
              0 & 0 & 0 & r\sin\theta \\
            \end{pmatrix},
\end{equation}
and $e=det(e^i_{\mu})=e^{\frac{\lambda+\nu}{2}} r^2 \sin \theta$
and correspondingly the torsion scalar and its derivative is obtained as,
\begin{eqnarray}
T(r)&=&\frac{2e^{-\lambda}}{r}\left(\nu'+\frac{1}{r}\right),\\
T'(r)&=&\frac{2e^{-\lambda}}{r}\left(\nu''-\frac{1}{r^2}\right)-T\left(\lambda'+\frac{1}{r}\right).
\end{eqnarray}
The Einstein-Maxwell field equation in $f(T)$ gravity are given by,
\begin{eqnarray}
4\pi \rho+E^2&=&\frac{f}{4}-\left[T-\frac{1}{r^2}-\frac{e^{-\lambda}}{r}(\lambda'+\nu')\right]\frac{f_T}{2}, \label{max1}\\
4\pi p_r-E^2&=&\left[T-\frac{1}{r^2}\right]\frac{f_T}{2}-\frac{f}{4},\\
4\pi p_t+E^2&=&\left[\frac{T}{2}+e^{-\lambda}\left\{\frac{\nu''}{2}+\left(\frac{\nu'}{4}+\frac{1}{2r}\right)(\nu'-\lambda')\right\}\right]\frac{f_T}{2}-\frac{f}{4}, \label{max2}\\
\text{and}~~~~\frac{\cot\theta}{2r^2}T'f_{TT}&=&0, \label{tor}\\
(r^{2}E)'&=&4\pi r^{2}\sigma(r) e^{\frac{\lambda}{2}}\label{e123},
\end{eqnarray}
 The equation (\ref{e123}) equivalently gives,
\begin{equation}
E(r)=\frac{1}{r^{2}}\int 4 \pi r^{2}\sigma e^{\frac{\lambda}{2}}dr.
\end{equation}

Here $\rho,~p$ and $E(r)$ represent the matter density, isotropic pressure and electric field respectively of the charged fluid sphere and `prime' denotes differentiation with respect to the radial co-ordinate `r'.





\section{Conformal Killing Equation}\label{sec3}
To relate the geometry with matter, a systematic approach is the use of conformal
symmetry under conformal killing vectors (CKV) and it is described by the following formula,
\begin{equation}\label{con}
L_\xi  g_{ik}=\psi  g_{ik},
\end{equation}
where $L$ is the Lie derivative operator and $\psi$  is the conformal factor.
The vector $\xi$ generates the conformal symmetry in such a way that
the metric g is conformally mapped onto itself along $\xi$. For $\psi = 0$ then Eq. (\ref{con}) provides the killing vector, for
$\psi =$ constant, Eq. (\ref{con}) gives a homothetic vector and if  $\psi = \psi(x,t)$ then it gives conformal vectors. One can note that for
$\psi = 0$, the underlying spacetime is asymptotically flat and it implies that the Weyl tensor will also vanish. So by studying the conformal killing vectors
one can get a more clear idea about the spacetime geometry.\par
Now eq.(\ref{con}) can be written as,
\begin{equation}
L_\xi g_{ik}=\xi_{i;k}+\xi_{k;i}=\psi g_{ik},
\end{equation}
For the line element (\ref{line}), the conformal killing equations are written as,
\begin{eqnarray}
\xi^{1}\nu'=\psi,\,
\xi^{4}=C_1,\,
\xi^{1}=\frac{\psi r}{2},\,
\xi^{1}\lambda'+2\xi^{1},_1=\psi
\end{eqnarray}
~~~~~Where `prime' and `comma' stand for the derivative and partial derivative with respect to `r' and $C_1$ is a constant.\\
The above equations yield,
\begin{eqnarray}
e^{\nu}&=&C_2^{2}r^{2}, \label{eq11}\\
e^{\lambda}&=&\left(\frac{C_3}{\psi}\right)^{2},\\
\xi^{i}&=&C_1\delta_{4}^{i}+\left( \frac{\psi r}{2}\right)\delta_1^{i}.
\end{eqnarray}
Where $C_2$ and $C_3$ are constants of integrations.\par

Now eqn.(\ref{tor}) yields,
\begin{equation}\label{1}
    f(T)=aT+b,
\end{equation}
where $a$ and $b$ are two constants of integration.\\
Using eqns. (\ref{eq11})-(\ref{1}) Einstein-Maxwell field equations (\ref{max1})-(\ref{max2}) becomes
\begin{eqnarray}
4\pi\rho &=& \frac{1}{4}\left(\frac{a}{r^2}+b\right)-\frac{3a}{2r}\frac{\psi\psi'}{C_3^2},\label{f1}\\
  4\pi p &=& \frac{a}{r^2}\frac{\psi^2}{C_3^2}+\frac{a\psi\psi'}{2rC_3^2}-\frac{1}{4}\left(\frac{a}{r^2}+b\right),\label{f2} \\
  E^2&=& -\frac{a}{2r^2}\frac{\psi^2}{C_3^2}+\frac{a\psi\psi'}{2rC_3^2}+\frac{a}{4r^2}. \label{f3}
  \end{eqnarray}

Now we have to solve the above field equations (\ref{f1})-(\ref{f3}) in three different regions of the charged gravastar which will be described one by one in the coming section.

\section{Solutions of the field equations in three different regions of the gravastar}\label{sec4}

\subsection{Interior Spacetime of the Charged Gravastar}
Now to obtain the metric coefficient for $e^{\lambda}$, let us consider the following equation of state (EoS) mentioned earlier between the matter density $\rho$ and the isotropic pressure $p$ as,
\begin{equation}\label{f4}
p=-\rho,
\end{equation}
as proposed by Mazur and Mottola\cite{maz,maz1}.\par
Now, using the relation between the matter density and isotropic pressure as given in eqn.(\ref{f4}), from eqns. (\ref{f1}) and (\ref{f2}) one can readily obtain a differential equation which is linear in conformal factor $\psi$ as follows:
\begin{equation}\label{eq1}
\frac{a\psi}{r^2C_3^2}(\psi-\psi'r)=0,
\end{equation}
From eqn. (\ref{eq1}) we get two solutions  for $\psi$. \[\text{either}~~\psi=0 ~~\text{or}~~ \psi=\psi_0r\]
where $\psi_0$ is the constant of integration. Since $\psi=0$ implies the asymptotically flat spacetime, we take $\psi=\psi_0r$ to calculate the other physical parameters in the interior region of the charged gravastar model.\\
Once we have obtained the expression of the conformal factor $\psi$, the expression for the metric coefficients in terms of the conformal factor can be written as,
\begin{eqnarray}
e^{\nu}&=&e^{-\lambda}={\tilde{\psi_0}}^2 r^2,
\end{eqnarray}
Here we have used the notation $\tilde{\psi_0}=\frac{\psi_0}{C_3}$ which is an another constant along with $C_2C_3=\psi_0$. One interesting thing is that here the metric coefficient $e^{\nu}$ is directly proportional to $r^2$ but $e^{\lambda}$ is inversely proportional to $r^2$.\\
 The expression for matter density $\rho$ and the isotropic pressure are obtained as,
\begin{eqnarray}
  4\pi\rho &=& \frac{1}{4}\left(\frac{a}{r^2}+b\right)-\frac{3}{2}a{\tilde{\psi_0}}^2=-4\pi p, \label{i1}
\end{eqnarray}
As it is well known that in the interior region of the gravastar, the energy density is positive, from eqn. (\ref{i1}) we get $\tilde{\psi_0}^{2}<\frac{1}{6a}\left(\frac{a}{r^2}+b\right)$ and therefore this inequality provides the upper bound of $\tilde{\psi_0}$. One can note that both the pressure and the density are inversely proportional to $r^{2}$ and they blow up without bound as we approach to the center of the gravastar, i.e., both suffer from central singularities and it is a natural behavior of the CKV model.\\
The expression for the electric field and charge density are readily obtained as,
\begin{eqnarray}
E^2&=& \frac{a}{4r^2} ,\label{e11}\\
\sigma(r)&=& \frac{\sqrt{a}\tilde{\psi_0}}{8\pi r}.
\end{eqnarray}
Charge $E$ is inversely proportional to $r$ and it does not depend on $\tilde{\psi_0}$. The charged density $\sigma$ depends on $\tilde{\psi_0}$ and inversely proportional to $r$.  \\
Next we obtain the active gravitational mass $M(r)$ as,
\begin{eqnarray}
M(r)=\int_0^{r} 4\pi \chi^{2}\left[\rho(\chi)+\frac{E^{2}(\chi)}{4\pi}\right]d\chi=\frac{br^3}{12}+\frac{ar}{2}(1-\tilde{\psi_0}^2r^2).
\end{eqnarray}
As $M(r)$ approaches to zero when $r\rightarrow 0$ , the active gravitational mass $M(r)$ does not face central singularity. Using the upper bound on $\tilde{\psi_0}$ as mentioned above, we get a lower bound of $M(r)$ as, $M(r)>\frac{5ar}{12}$ and at the same time it is clear that gravitational mass function is positive inside the interior of the charged gravastar since eqn. (\ref{e11}) verifies that `a' is positive.

\subsection{Thin shell of the charged gravastar}
In the shell of the gravastar, following the concept of Mazur \& Mottola \cite{maz}, the relation between the pressure $p$ and the energy density $\rho$ is taken as,
\begin{equation}\label{e2}
p=\rho ,
\end{equation}
This EoS is a special case of barotropic EoS $p=\gamma \rho$ with $\gamma=1.$ In one of our previous papers \cite{fr1}, we used the barotropic EoS to obtain a new class of exact solutions for the interior in $(2 + 1)$-dimensional spacetime for the perfect fluid model both with and without cosmological constant $\Lambda$.  \\
From eqns. (\ref{f1}) and (\ref{f2}) with the help of eqn. (\ref{e2}) we get the following ODE,
\begin{equation}
\frac{\psi^2}{C_3^2}+\frac{2r\psi\psi'}{C_3^2}=\frac{1}{2}\left(1+\frac{br^2}{a}\right),
\end{equation}
Solving the above equation we obtain,
\begin{eqnarray}
  \frac{\psi^{2}}{C_3^2} &=& \frac{1}{2}+\frac{b}{6a}r^2-\frac{\tilde{\psi_1}}{r},
  \end{eqnarray}
  Here $\tilde{\psi_1}$ is a constant of integration.\\ The coefficients of $g_{rr}$ and $g_{tt}$ are obtained as,
  \begin{eqnarray}
   e^{-\lambda} &=& \frac{1}{2}+\frac{b}{6a}r^2-\frac{\tilde{\psi_1}}{r},\\
  e^{\nu}&=&C_2^2r^2.
  \end{eqnarray}
  The matter density and isotropic pressure inside the thin shell are obtained as,
  \begin{eqnarray}
  4 \pi \rho &=& \frac{a}{4r^2}-\frac{3a\tilde{\psi_1}}{4r^3}=4 \pi p ,
  \end{eqnarray}
  The electric field and electric charge density inside the thin shell take the form
  \begin{eqnarray}
  E^2&=& \frac{3a\tilde{\psi_1}}{4r^3} , \label{e}\\
  \sigma(r)&=& \frac{1}{16\pi r^2}\sqrt{\frac{3a\tilde{\psi_1}}{r}}\sqrt{\frac{1}{2}+\frac{br^2}{6a}-\frac{\tilde{\psi_1}}{r}}. \label{sig0}
\end{eqnarray}
Eqn.(\ref{e}) confirms that $\tilde{\psi_1}$ is positive. We note that in the interior region of the gravastar $E$ does not depend on $\tilde{\psi_1}$ but in the present case electric field $E$ depends on $\tilde{\psi_1}$. Here $E$ is inversely proportional to $r^{3/2}$ and both the pressure and density suffer from central singularity like the previous case. Eqn. (\ref{sig0}) implies that $\tilde{\psi_1}<\frac{(3a+br^2)r}{6a}$, so combining the above two cases we get a bound for $\tilde{\psi_1}$ as, $0<\tilde{\psi_1}<\frac{(3a+br^2)r}{6a}$.

\subsection{Exterior Spacetime and junction condition}
In the exterior spacetime of the charged gravastar the relationship between the pressure and density is $ p=\rho=0 $ and the exterior spacetime is described by the  Reissner-Nordstr\"{o}m line element given by,
\begin{eqnarray}
ds^{2}&=&\left(1-\frac{2m}{r}+\frac{q^2}{r^{2}}\right)dt^{2}-\left(1-\frac{2m}{r}+\frac{q^2}{r^{2}}\right)^{-1}dr^{2}-r^{2}(d\theta^{2}+\sin^{2}\theta d\phi^{2}),
\end{eqnarray}
where, $r>m+\sqrt{m^{2}-q^2}$, $m$ being the mass and $q$ being the charge of the black hole.
So we can match our interior spacetime to the exterior R-N spacetime at the boundary of the charged gravastar.\par

Now to determine the surface stresses at the junction boundary we use the Darmois-Israel \cite{dm1} formation. The intrinsic surface stress energy tensor $S_{ij}$ is given by Lancozs equations in the following form
\begin{equation}
S^{i}_{j}=-\frac{1}{8\pi}(\kappa^{i}_j-\delta^{i}_j\kappa^{k}_k),
\end{equation}
where $\kappa_{ij}$ represents the discontinuity in the second fundamental form is written as,
\begin{equation}
\kappa_{ij}=K_{ij}^{+}-K_{ij}^{-},
\end{equation}
$K_{ij}$ being the second fundamental form is presented by
\begin{equation}
K_{ij}^{\pm}=-n_{\nu}^{\pm}\left[\frac{\partial^{2}X_{\nu}}{\partial \xi^{i}\partial\xi^{j}}+\Gamma_{\alpha\beta}^{\nu}\frac{\partial X^{\alpha}}{\partial \xi^{i}}\frac{\partial X^{\beta}}{\partial \xi^{j}} \right]|_S,
\end{equation}
where $n_{\nu}^{\pm}$ are the unit normal vectors defined by,
\begin{equation}
n_{\nu}^{\pm}=\pm\left|g^{\alpha\beta}\frac{\partial f}{\partial X^{\alpha}}\frac{\partial f}{\partial X^{\beta}}  \right|^{-\frac{1}{2}}\frac{\partial f}{\partial X^{\nu}},
\end{equation}
with $n^{\nu}n_{\nu}=1$. Where $\xi^{i}$ is the intrinsic coordinate on the shell. $+$ and $-$ corresponds to exterior i.e., R-N spacetime and interior spacetime respectively.\par
Using the spherical symmetry nature of the spacetime, the surface stress energy tensor can be written as $S^{i}_j=diag(-\Sigma,\mathcal{P})$. Where $\Sigma$ and $\mathcal{P}$ being the surface energy density and surface pressure respectively.\\
The expression for surface energy density $\Sigma$ and the surface pressure $\mathcal{P}$ at the junction surface $r = r_b$ are obtained as,
\begin{eqnarray}
\Sigma &=&-\frac{1}{4\pi r_{b}}\left[\sqrt{e^{-\lambda}} \right]_{-}^{+}=-\frac{1}{4\pi r_b}\left[\sqrt{1-\frac{2M}{r_b}+\frac{q^2}{r_b^2}}-\tilde{\psi_0}r_b\right],\label{sig}\\
\mathcal{P}&=&\frac{1}{8\pi r_b}\left[\left\{1+\frac{a\nu'}{2}\right\}\sqrt{e^{-\lambda}}\right]_{-}^{+}
=\frac{1}{8\pi r_b}\left[\frac{1-\frac{M}{r_{\Sigma}}}{\sqrt{1-\frac{2M}{r_{\Sigma}}+\frac{q^2}{r_b^2}}}-2\tilde{\psi_0}r_b\right].
\end{eqnarray}

Hence we have matched our interior charged gravastar solution to the exterior Reissner-Nordstr\"{o}m spacetime in presence of thin shell.

Now we are interested to find the mass of the thin shell ($m_s$) and it can be obtained from the following formula:
\begin{equation}\label{ms}
m_s=4\pi r_b^2\Sigma,
\end{equation}
Employing the expression of $\Sigma$ given in (\ref{sig}) into eqn. (\ref{ms}) and rearranging, the mass of the charged gravastar in terms of the thin shell mass is calculated as,
\begin{equation}
m=\frac{r_b}{2}\left[1+\frac{q^2}{r_b^2}-\frac{m_s^2}{r_b^2}-\tilde{\psi_0}^{2}r_b^2+2m_s\tilde{\psi_0}\right].
\end{equation}
\section{Some physical properties}\label{sec5}
We can obtain the proper thickness of the shell that joins the interior region of the charged gravastar with the exterior region by the following formula :
\begin{eqnarray}\label{l0}
  \ell &=& \int_{r_1}^{r_2}\sqrt{e^{\lambda}} dr=\int_{r_1}^{r_2}\frac{1}{\sqrt{\frac{1}{2}+\frac{br^2}{6a}-\frac{\psi_1}{r}}}dr,
\end{eqnarray}
 As the thickness of the shell being very small, to perform the integral we set $r_1=d$ and $r_2=d+\epsilon$, and it provides,
\begin{equation}\label{l}
\ell=\int_{d}^{d+\epsilon}\frac{1}{\sqrt{G(r)}}dr,
\end{equation}
where $G(r)=\frac{1}{2}+\frac{br^2}{6a}-\frac{\psi_1}{r}$. Now it is very difficult to perform the integral given in eqn.(\ref{l}). let $F(r)$ be the primitive of $\frac{1}{\sqrt{G(r)}}$. Now by using the fundamental theorem of integral calculus, from equation (\ref{l}) we obtain,
\begin{equation}\label{l1}
\ell=\left[F(r)\right]_{d}^{d+\epsilon}=F(d+\epsilon)-F(d),
\end{equation}
Expanding $F(d+\epsilon)$ in Taylor series about $`d'$ and retaining up to the linear order of $\epsilon$, from eqn. (\ref{l1}) we obtain
\begin{equation}\label{l2}
\ell=\frac{\epsilon}{\sqrt{\frac{1}{2}+\frac{bd^2}{6a}-\frac{\psi_1}{d}}},
\end{equation}
Let us now calculate the energy $\tilde{E}$ within the shell from the following formula
\begin{equation}\label{l3}
\tilde{E}=\int_{d}^{d+\epsilon}4\pi r^2\left(\rho+\frac{E^2}{4\pi}\right)dr=\frac{a\epsilon}{4},
\end{equation}
Following the concept of Mazur and M\"{o}ttola \cite{maz}, we calculate the entropy within the shell of the gravastar as,
\begin{equation}\label{l4}
S=\int_{d}^{d+\epsilon}4\pi r^2 s(r)\sqrt{e^{\lambda}}dr,
\end{equation}
$s(r)$ being the entropy density for the local temperature $T(r)$ and is defined by, $$s(r)=\alpha \left(\frac{K_B}{\hbar}\right)\sqrt{\frac{p}{2\pi}}$$ Where $\alpha^{2}$ is a dimensionless constant.
Using the expression for $p$ and $e^{\lambda}$, from eqn.(\ref{l4}), we calculate the expression for entropy as follows:
\begin{equation}\label{l5}
S(r)=\int_d^{d+\epsilon} a \alpha \left(\frac{K_B}{\hbar}\right)r \sqrt{\frac{3(r-3\psi_1)}{3ar+br^3-6a\psi_1}},
\end{equation}
To find the expression of $S(r)$ we have to perform the integral of eqn (\ref{l5}). But it is difficult to perform the integration. Using the same approach as we did for our previous integral of eqn. (\ref{l0}), we obtain,
\begin{equation}\label{s}
S(r)= d \epsilon a \alpha \left(\frac{K_B}{\hbar}\right)\sqrt{\frac{3(d-3\psi_1)}{3ad+bd^3-6a\psi_1}}.
\end{equation}
Hence the expression of the entropy is obtained for our proposed model.

\section{Discussion}\label{sec6}
Present paper deals with the formation of analytic model of gravastar in $f(T)$ gravity with charged matter distribution. The interior spacetime is taken as static charged isotropic spherical source having three different regions with three different equation of state. With the help of the diagonal tetrad field, the Einstein-Maxwell field equations in $f(T)$ gravity have been formulated to obtain the gravastar model. Eqn.(\ref{tor}) implies that unknown function $f(T)$ appears as a linear function of $T$ i.e., $f(T) = aT+b$, where $a$ and $b$ are the constants of integration. Taking this form of $f(T)$ with the three different equations of state, we have solved the field equations and obtained explicitly the expression for $e^{\lambda}$, matter distribution, isotropic pressure, electric field and charged density in the interior region as well as in the thin shell. In the interior region of the gravastar the electric field does not depend on $\tilde{\psi_0}$ but in case of the thin shell of the gravastar the electric field depends on $\tilde{\psi_1}$. We obtained the bounds for both the integration constants $\tilde{\psi_0}$ and $\tilde{\psi_1}$. In the interior region of the gravastar both the pressure and density are inversely proportional to $r^{2}$. In present model the matter density, isotropic pressure, electric field and electric charge density all blows up without bounds if one approaches to the center of the gravastar i.e., they suffers from central singularity as it is a common behavior of the CKV model. One can also notice that all the physical parameters e.g.,~length of the thin shell, energy, entropy etc. match to the result of Usmani et al. \cite{usmani} for $a=1$ and $b=0$  i.e., our model come backs to the model of GTR. The expression for effective gravitational mass has also been worked out and it is regular at center and positive inside the interior region of the gravastar. We have calculated the mass of the thin shell and also expressed the mass of the charged gravastar in terms of the mass of the thin shell. The surface energy density and surface pressure are also calculated at the junction boundary by using Darmois-Israel \cite{dm1} formation.\par
Now a general question arises in our mind that what are the possible observational signatures for this kind of gravstar? Though there are no direct evidences to detect garavastar but some of the indirect ways have been discussed in literature. The idea for possible detection of gravastar was first proposed by Sakai et al. \cite{sakai} through the study of gravastar shadows. They investigated the optical
images of the gravastars possessing unstable circular orbits
of photons, assuming its optically transparent surface and
two types of optical sources behind a gravastar: an infinite
optical plane and a companion star. Broderick and Narayan \cite{bn} discussed observational constraints on gravastar models with the thermal process. Pani et al. \cite{pani} discussed how the presence or absence of an event horizon can produce qualitative differences in the gravitational waves emitted by ultracompact objects. Uchikata et al. \cite{u} showed how a measurement of the tidal deformability from the gravitational-wave detection of a compact-binary inspiral can be used to constrain gravastars. Kubo and Sakai \cite{kubo} proposed a possible method to detect gravastars, by studying their gravitational lensing effects. They also calculated the image of a companion which rotates around the gravastar and found that some characteristic images appear depending on whether the gravastar possess unstable circular orbits of photons or not. Cardoso et al. \cite{car1,car2} showed that only late-time ringdown detections might be used to rule out exotic alternatives to BHs and to test quantum effects at the horizon scale. However, we believe that our approach provides a useful tool to describe a model of charged gravstar in $f(T)$ modified theory of gravitation.

\bibliographystyle{unsrt}
\bibliography{grav_final}

\end{document}